\documentclass[onecollarge,natbib]{svjour2}
\bibpunct{[}{]}{;}{n}{}{,} % to get "[numbered]" references from natbib
\smartqed  % flush right qed marks, e.g. at end of proof
\usepackage{amsmath,revsymb,graphicx,amssymb,array}
\newcommand{\kt}[1]{\ensuremath{|#1\rangle}}

\newcommand{\bk}[2]{\ensuremath{\langle #1|#2\rangle}}
\journalname{Few-Body Systems}
\begin{document}

\title{Identical Wells, Symmetry Breaking, and the Near-Unitary Limit%\thanks{Grants or other notes
}


\author{N.L. Harshman
}


\institute{N.L. Harshman \at
              Department of Physics, American University, 4400 Massachusetts Ave.\ NW, Washington, DC 20016-8058, USA \\
              \email{harshman@american.edu}             \\
           \emph{Present address:} Department of Physics and Astronomy, Aarhus University, Ny Munkegade 120, DK-8000 Aarhus C  
}

\date{Received: date / Accepted: date}

\maketitle

\begin{abstract}
Energy level splitting from the unitary limit of contact interactions to the near unitary limit for a few identical atoms in an effectively one-dimensional well can be understood as an example of symmetry breaking. At the unitary limit in addition to particle permutation symmetry there is a larger symmetry corresponding to exchanging the $N!$ possible orderings of $N$ particles. In the near unitary limit, this larger symmetry is broken, and different shapes of traps break the symmetry to different degrees. This brief note exploits these symmetries to present a useful, geometric analogy with graph theory and build an algebraic framework for calculating energy splitting in the near unitary limit.
\end{abstract}

\section{Introduction}

The primary motivation for this work is the problem of a few strongly-interacting identical particles in one-dimension. Inspired by many-body~\cite{kinoshita_2004, paredes_2004, kinoshita_2006, haller_2009} and few-body experiments~\cite{serwane_deterministic_2011, wenz_few_2013, murmann_two_2015, murmann_antiferromagnetic_2015} with ultracold atoms in optical traps interacting via tunable Feshbach and confinement resonances, the following Hamiltonian has recently received a lot of attention:
\begin{equation}\label{eq:ham0}
H = \sum_{i=1}^N \left( - \frac{\hbar^2}{2 m} \frac{\partial^2}{\partial x_i^2} + V(x_i) \right) + g \sum_{\langle i,j \rangle}  \delta(x_i - x_j).
\end{equation}
This Hamiltonian models $N$ particles in an effectively one-dimensional trap with shape $V(x)$ with zero-range, contact interactions~\cite{olshanii_1998}. The limit $g\to\infty$ is called the hard-core or unitary limit, and for any trap $V(x)$, the stationary states for identical fermions and boson with and without spin or internal components can be constructed by generalizations~\cite{Girardeau2007, Deuretzbacher2008, Yang2009, Guan2009, Ma2009,  Girardeau2010a, Girardeau2010b, Girardeau2011, Fang2011, Cui2014, Harshman2016b} of the famous Girardeau Fermi-Bose mapping~\cite{Girardeau1960}. Given the one-particle trap eigenstates, these solutions are exact and the system is integrable. 

While experiments can approach this limit, the near-unitary limit of (\ref{eq:ham0}) is necessary to interpret actual experiments and to construct maps that connect the non-interacting $g=0$ solutions and the Girardeau-like solutions in the unitary limit under adiabatic tuning of the parameter $g$. A series of papers~\cite{Deuretzbacher2014, Volosniev2013, Levinsen2014, Gharashi2015, Yang2015, Decamp2016} have investigated the near-unitary limit by a variety of approaches, including first- and second-order perturbation theory, analytic ansatz, and numerical methods. A key result is that in the near-unitary limit the system can be reduced to a spin chain model whose site-coupling coefficients depend on the specific shape of the trap. Such models can be analytically solved for any $N$ and any number of spin components, and there are now numerical packages available for the solution of the required coupling coefficients~\cite{CONAN}.

This brief note contributes to the literature on the unitary and near-unitary limit in two ways. First, it provides a geometrical picture that clarifies the role of symmetry and allows the near-unitary limit to be understood via the language symmetry breaking. In the process, it provides an algebraic framework that allows separation of generic properties of solutions due to symmetry from contingent properties depending on the details of the trap shape. Second, it distinguishes trio of distinct symmetries and demonstrates their usefulness: well permutation symmetry, ordering permutation, and particle permutation symmetry. The author has argued elsewhere~\cite{harshman2016c} that exploiting the full, interrelated structure of these symmetries shows promise in analyzing the next generation of few-atom, few-well experiments.

\section{One particle in six identical wells}

Begin by considering an alternate system which will prove to be analogous: one particle in a trapping potential with six identical, isolated wells. To simplify the model, assume that the the energy spectrum for each individual well is non-degenerate and denote the single-well energies by $\epsilon_n$ for $n\in\mathbb{N}$. Labeling the wells with $W\in\{A,B,C,D,E,F\}$, the Hamiltonian can be written as 
\begin{equation}\label{eq:ham1}
\hat{H}^{(0)} = \hat{H}_A +\hat{H}_B +\hat{H}_C +\hat{H}_D +\hat{H}_E + \hat{H}_F
\end{equation}
where all the sub-Hamiltonians $\hat{H}_W$ commute because they are defined on disjoint regions of space. The Hamiltonian $\hat{H}^{(0)}$ is symmetric under permutations of the six wells. We can think of well permutations as active (physically exchanging the locations of the wells) or passively (exchanging the labels of the wells), but either way a well permutation just rearranges sub-Hamiltonians in the sum (\ref{eq:ham1}). 

A natural basis of energy eigenstates is provided by vectors $\kt{W_n}$ where $n\in\mathbb{N}$ labels the single-well energy. These vectors have the properties $\bk{W_n}{W'_N} = \delta_{WW'}\delta_{nn'}$ and $\hat{H}^{(0)}\kt{W_n} = \epsilon_n\kt{W_n}$. Define the well permutation operators $\hat{P}(p)$, where $p$ is a permutation in the symmetric group of six objects $\mathrm{S}_6$. Using cycle notation for $p$, the element $p=(WW')$ is represented by the operator $\hat{P}(WW')$ exchanges wells $W$ and $W'$. They act on the basis $\kt{W_n}$ as one one would expect: $\hat{P}(AB)\kt{A_n} = \kt{B_n}$, $\hat{P}(AB)\kt{B_n} = \kt{A_n}$, and $\hat{P}(AB)\kt{C_n} = \kt{C_n}$, for example. 

A note for people interested in representation theory (this may be skipped without losing the thread of the argument): The restriction of $\hat{P}(p)$ to any energy eigenspace is a $6\times 6$ matrix of ones and zeros and is called the `natural' or `defining' of the symmetric group $\mathrm{S}_6$. It is not an irreducible representation (irrep), but it can be reduced into one copy of the totally symmetric irrep  of $\mathrm{S}_6$ and one copy of the standard irrep of $\mathrm{S}_6$. The fact that energy eigenspaces are not irreducible under $\mathrm{S}_6$ means that Hamiltonian $\hat{H}^{(0)}$ must have another symmetry. This symmetry is provided by the uncoupled dynamics of the six well. When the wells are truly isolated, the time evolution operator generated by each sub-Hamiltonian forms an independent realization of the one-parameter Lie group of time translations $\mathrm{T}_t$. The full symmetry group of $\hat{H}^{(0)}$ is therefore $\mathrm{S}_6 \ltimes (\mathrm{T}_t)^{\times 6}$, where  there are six copies of $\mathrm{T}_t$ and the semidirect product $\ltimes$ means that $\mathrm{S}_6$ acts on $(\mathrm{T}_t)^{\times 6}$  by permutations `naturally'. This group is also denoted by the wreath product $\mathrm{S}_6 \wr \mathrm{T}_t$ and this group has a six-dimensional irrep with the right properties to realize energy eigenspaces of $\hat{H}^{(0)}$; it also has other irreps that are not useful in this context. For more details see~\cite{Harshman2016b, harshman2016c, harshman2016a}.

To return to the main point, every energy level of $\hat{H}^{(0)}$ is six-fold degenerate when the wells are perfectly isolated (again, assuming that each well has a non-degenerate spectrum; otherwise there are additional degeneracies). However, when tunneling is present, this degeneracy will be partially or totally broken depending on the details of the tunneling. Specifically, I will consider that the tunneling between two wells depends on the distance between them and then use symmetry analysis to build a tunneling operator to calculate the splitting of the six-fold degeneracy at first order.

\begin{figure*}
\centering
  \includegraphics[width=0.95\textwidth]{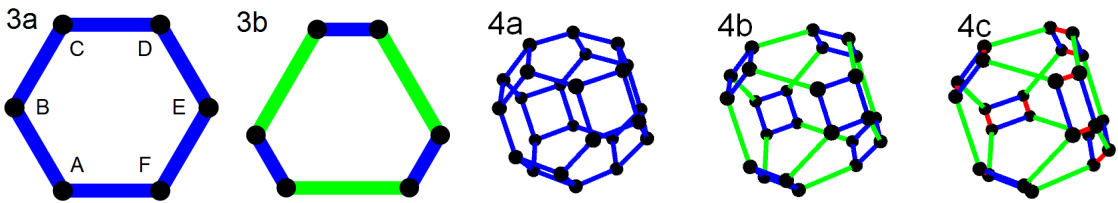}
\caption{These five figures depict graphs that represent several possibilities for symmetry breaking by tunneling for six wells and twenty-four wells in configurations that are relevant to the analysis of  the near-unitary limit. In the six-well figures, each vertex corresponds to a particular order of three particles and the the twenty-four well figures each vertex corresponds to an ordering of four particles. The colored lines connect vertices for which tunneling is allowed at the lowest order of perturbation theory; the length of the line represents the strength of the tunneling. Subfigure 3a represents equal tunneling rate for the first two particles exchanging and for the second two particles exchanging; for subfigure 3b these rates are different. Subfigures 4a, 4b, and 4c correspond to all three pairwise exchanges having the same tunneling rate; to the first two and the last two having the same rate but different from the middle pair; and to all three pairwise exchanges having different rates. Notice that subfigures 3a and 3b both have at least $\mathrm{D}_3 \sim \mathrm{S}_3$ symmetry, realizing particle permutations. Subfigures 4a-4c all have at least $\mathrm{T} \sim \mathrm{S}_4$ symmetry.}
\label{fig:AAA}       % Give a unique label
\end{figure*}

Figure \ref{fig:AAA} depicts two configurations of the six wells. In the first configuration 3a, each well is equidistant from two neighbors and tunneling between anything but the nearest two neighbors will be neglected. In the second configuration 3b, each well is closer to one neighbor than the other and the analysis will account for different tunneling rates to each neighbor. In either configuration, the $\mathrm{S}_6$ symmetry of isolated wells is broken (and so is the larger symmetry group $\mathrm{S}_6 \wr \mathrm{T}_t$). In the first configuration, the remaining symmetry is isomorphic to dihedral group $\mathrm{D}_6$ with order 12, the point symmetries of a hexagon. In the second configuration has the dihedral symmetry $\mathrm{D}_3$ with order 6, the point symmetries of a triangle.

For both cases, and for each energy eigenspace of $\hat{H}^{(0)}$ with energy $\epsilon_n$, we can construct a tunneling operators that will give the correct energy splittings and energy eigenstates for the first order perturbation. The two-particle permutation operators like $\hat{P}(WW')$ provide a matrix element between wells where tunneling is allowed. For the first configuration 3a, the operator is
\begin{equation}\label{eq:t1}
\hat{T}_n = -t_n \left( \hat{P}(AB) + \hat{P}(BC) + \hat{P}(CD) + \hat{P}(DE) + \hat{P}(EF) + \hat{P}(FA) \right)
\end{equation}
The coefficients $t_n$ have units of energy and measure the rate of the tunneling at first order. Generally, they depend on the energy level under consideration and cannot be calculated from the isolated-well Hamiltonian $\hat{H}^{(0)}$ without more information about the boundary region in which the six nearly-isolated wells are embedded. The eigenvalues of (\ref{eq:t1}) are in increasing order $\{-6 t_n, -5 t_n,  -3 t_n,  -2t_n\}$ with corresponding orthonormalized eigenvectors of $\hat{T}_n$ in the $\epsilon_n$ subspace are
\begin{eqnarray*}
-6 t_n &\to& \frac{1}{\sqrt{6}} \left(\kt{A_n}+\kt{B_n}+\kt{C_n}+\kt{D_n}+\kt{E_n}+\kt{F_n}\right)\\
-5 t_n &\to& \left\{\begin{array}{l}
\frac{1}{2} \left(\kt{A_n}- \kt{C_n} - \kt{D_n} +\kt{F_n}\right)\\
\frac{1}{2\sqrt{3}} \left(\kt{A_n}+2\kt{B_n}+\kt{C_n}-\kt{D_n}-2\kt{E_n}-\kt{F_n}\right)
\end{array} \right.\\
-3 t_n &\to& \left\{\begin{array}{l} \frac{1}{2} \left(\kt{A_n}- \kt{C_n} + \kt{D_n} -\kt{F_n}\right)\\
\frac{1}{2\sqrt{3}} \left(\kt{A_n}-2\kt{B_n}+\kt{C_n}+\kt{D_n}-2\kt{E_n}+\kt{F_n}\right) \end{array} \right.\\
-2 t_n &=& \frac{1}{\sqrt{6}} \left(\kt{A_n}-\kt{B_n}+\kt{C_n}-\kt{D_n}+\kt{E_n}-\kt{F_n}\right).
\end{eqnarray*}
The two non-degenerate states are the totally symmetric irrep (with greatest energy shift) and the totally antisymmetric irrep of $\mathrm{D}_6$ (with least shift). Notice that even after the splitting from tunneling, there are two two-fold degenerate states corresponding to two-dimensional irreps of $\mathrm{D}_6$. One way to distinguish them physically it to look at inversion around the origin (equivalently, rotating by $\pi$). This operator can be constructed from well exchanges as $\hat{\Pi}=\hat{P}(AD)\hat{P}(BE)\hat{P}(CF)$. The states with energy shifts $-6t_n$ and $-3t_n$ are parity symmetric, and the other three states are odd under parity.

To be clear, this tunneling operator (\ref{eq:t1}) is only giving the relative splitting in the energy. For example, the spread in energies is $4 t_n$ with the specific pattern above is a generic property of the unitary limit with symmetric coupling. For fully-connected graphs like 3a and 3b, to find the actual energy one can add an additional operator that is proportional to the identity, and therefore commuting with $\hat{H}^{(0)}$ and $\hat{T}_n$. More detailed information about the boundary region is required to fix the energy parameter multiplying this identity operator.

For the second configuration 3b the tunneling operator is
\begin{equation}\label{eq:t2}
\hat{T}'_n = -t_n \left( \hat{P}(AB)  + \hat{P}(CD)  + \hat{P}(EF)\right) - u_n \left(\hat{P}(BC) + \hat{P}(DE) + \hat{P}(FA) \right)
\end{equation}
with eigenvalues $\{-3(t_n+u_n), -2(t_n + u_n) \pm \sqrt{t_n^2 -t_nu_n + u_n^2}, -(t_n +u_n)\}$. 
As before the totally symmetric irrep of the symmetry group (now $\mathrm{D}_3$) has the greatest shift and the totally antisymmetric has the least shift. Parity is no longer a symmetry and cannot be used to distinguish the two two-fold degenerates states with energies $-2(t_n + u_n) \pm \sqrt{t_n^2 -t_nu_n + u_n^2}$ which transform under the same irrep of $\mathrm{D}_3$.

\section{Near-unitary limit of three particles in one dimension}

The essential argument is that the Hamiltonian for three particles in one dimension that have infinitely hard-core interactions (including finite range and not just contact interactions) can be put into the form (\ref{eq:ham1}) and that the near-unitary limit can be calculated for symmetric traps using (\ref{eq:t1}) and for asymmetric traps by (\ref{eq:t2}). The key idea is that when there are hard-core interactions, the configuration space $\mathbb{R}^3$ is broken into six distinct domains $\langle ijk \rangle \equiv \{ (x_1, x_2, x_3) \in \mathbb{R}^3 | \ x_i < x_j < x_k \} $. These six domains are all identical, and their dynamics are independent in the unitary limit, just like the previous example. The near-unitary limit, where there is a small chance for particles to exchange, can be thought of as tunneling among adjacent domains. The rest of this section elucidates this analogy and then provides additional details for contact interactions in the unitary limit.

The six wells of the previous section now correspond to the six orderings of three particles in one dimension. Choose the map between orderings and wells such that
\begin{equation}
A\rightarrow \langle 123 \rangle,\ B\rightarrow \langle 132 \rangle,\ C\rightarrow \langle 312 \rangle,\ D\rightarrow \langle 321 \rangle,\ E\rightarrow \langle 231 \rangle,\ \mbox{and}\ F\rightarrow \langle 213 \rangle.
\end{equation}
With this assignment, the wells of Fig.\ 1.3a correspond schematically to the locations of the domains $\langle ijk \rangle$ in relative configuration space using standard Jacobi coordinates $x \propto x_1 - x_2$ and $y \propto x_1 + x_2 - 2 x_3$. In the unitary limit there are two useful subgroups of the \emph{well permutation} symmetry  $\mathrm{S}_6$.
\begin{itemize}
\item The first is the subgroup of \emph{particle permutations} which permute the particles no matter where they are in the order $\langle ijk \rangle$. For example, exchanging particles 1 and 2 exchanges domains $A \leftrightarrow F$, $B \leftrightarrow E$ and $C \leftrightarrow D$. In configuration space, particle permutations are linear transformations. In particular, exchanging particles 1 and 2 is a reflection across the plane $x_1 - x_2 =0$ and in Fig.~1.3a this is reflection across the vertical. This transformation can be written in terms of the well permutation operators as $\hat{P}(AF)\hat{P}(BE)\hat{P}(CD)$. The subgroup of particle permutations remain as a symmetry in the near unitary limit.
\item The second is the subgroup of \emph{ordering permutations}. These permute the positions, no matter which particle is in that position. For example, exchanging the left-most and the center particle so that $\langle ijk \rangle \to \langle jik \rangle$ exchanges domains $A \leftrightarrow F$, $B \leftrightarrow C$ and $D \leftrightarrow E$. This transformation can be written in terms of the well permutation operators as $\hat{P}(AF)\hat{P}(BC)\hat{P}(DE)$. Note that this is a non-linear transformation (or perhaps more descriptively, piece-wise linear) in configuration space and is only an exact symmetry in the unitary limit when the wave function is forced to have nodes at the boundaries between regions.
\end{itemize}
Ordering permutations and particle permutations are both isomorphic to $\mathrm{S}_3$ and they share no element except the identity. Remarkably, the well permutation subgroups of particle permutations and ordering permutations commute with each other. Each of the six degenerate states at the unitary limit can be jointly labeled by pair of $\mathrm{S}_3$ tableaux (see \cite{Harshman2016b,harshman2016a} for more details on the double tableaux basis).

For our purposes, the key property is that in the near-unitary limit, tunneling breaks ordering permutation symmetry but preserves particle permutation symmetry. The tunneling operator in (\ref{eq:t1}) corresponds to symmetric tunneling between the first and second particle and the second and third particle. This would be expected in symmetric one-dimensional traps, and parity inversion is realized by the operator $\hat{\Pi}$ defined in the last section. As expected, the totally symmetric state has the largest energy shift. The two mixed symmetry states have smaller shifts, and then finally the totally antisymmetric state has the smallest shift. 

If the particles are indistinguishable, then only states with the correct symmetry under particle permutations can be populated. For example, one-component bosons can only be in the totally symmetric state (either at the unitary or near-unitary limit), and one-component (e.g.\ polarized) fermions can only be in the totally antisymmetric state. If there are spin or other internal degrees of freedom to `carry' the necessary symmetry or antisymmetry, then other states can be populated.

When the trap is not symmetric, then the tunneling parameter for exchanges of the first and second particle need not be the same as the second and third particle, and so the second tunneling operator (\ref{eq:t2}) is appropriate. There is no longer parity symmetry, and now the splitting depends on two parameters. However the qualitative structure is the same as before.

In the specific case of contact interactions at the unitary limit, then much more can be said. Each six-fold degenerate energy level can be built by constructing the totally antisymmetric combination of three non-interacting single-particle state:
\begin{equation}
\Phi_n(x_1,x_2,x_3) = \sum_{\langle ijk \rangle} (-1)^{\pi\langle ijk \rangle} \phi_{n_i}(x_1)\phi_{n_j}(x_2)\phi_{n_k}(x_3)
\end{equation}
where $\phi_{n_i}(x)$ is a single-particle trap eigenstate, the sum is over all permutations of ${\langle 123 \rangle}$, and $\pi\langle ijk \rangle$ is the sign of the permutation. The label $n$ now stands for a triple of single particle energy quantum numbers, with the ground state $n=0$ corresponding to the set $\{012\}$, the first excited state $n=1$ to $\{013\}$ and the rest of the assignments depending on the one-dimensional trapping potential. Restricting $\Phi(x_1,x_2,x_3)$ to each of the six ordering domains $\langle ijk \rangle$ gives the states $\kt{W_n}$, which can be used to construct the first-order eigenstates as above. The values for $t_n$ (and, if necessary $u_n$) can be explicitly calculated using
\begin{eqnarray}
t_n &=& \frac{6 \hbar^4}{m^2 g} \int_{-\infty}^{+\infty} dx_3 \, \int_{-\infty}^{x_3} dx_2  \left|\frac{\partial \Phi_n(x_1,x_2,x_3)}{\partial x_1}\right|^2_{x_1=x_2}\nonumber\\
u_n &=& \frac{6 \hbar^4}{m^2 g} \int_{-\infty}^{+\infty} dx_1 \, \int_{-\infty}^{x_1} dx_2  \left|\frac{\partial \Phi_n(x_1,x_2,x_3)}{\partial x_2}\right|^2_{x_2=x_3}.
\end{eqnarray}
where $g$ is the large but finite strength of the contact interaction with units of energy times length~\cite{Deuretzbacher2014}. The additional identity operator needed to correctly capture the non-shift of the totally antisymmetric energy level is $t_n+u_n$, and for symmetric traps $t_n=u_n$.

\section{Four particles and other extensions}

This algebraic framework can be extended to more particles, although geometric analogies become less and less intuitive. Consider the last three subfigures of Figure \ref{fig:AAA} with twenty-four wells corresponding to all the possible orderings of four particles. Now there are three relevant cases depending on the relative tunneling rates for the first two, middle two and last two pairs. Subfigure 4a corresponds to all tunneling rates equal, as would occur in an ideal infinite square well trap. Subfigure 4b corresponds to a generic symmetric trap, and subfigure 4c to a generic non-symmetric trap. These three cases are listed in term of increasing breaking of well permutation symmetry. Although somewhat more cumbersome, tunneling operators can be constructed using computer algebra programs for $N=4$ and higher. 

More generally, these methods can be used to look at symmetry breaking in a variety of problem where in some limit the configuration space is partitioned into disjoint regions. It can be extended to wells with additional symmetries and degeneracies, and used to analyze integrability, separability, and solvability. In \cite{harshman2016c} these techniques are used to look at a two and more interacting particles in variable double-wells and construct adiabatic maps among limiting cases. These methods should be useful in the analysis of other experiments with a few particles in a few wells, such as \cite{kaufman2014, kaufman2015}, which promise to explore the `bottom-up' approach to condensed matter starting from a few cold atoms with tunable interactions in a few controllable wells.

\begin{acknowledgements}
The author would like to thank the support of the Aarhus University Research Foundation the hospitality of Nikolaj Zinner, Dmitri Fedorov and other Aarhus colleagues while this work was completed.
\end{acknowledgements}

\end{document}